\documentclass[aps,prl,reprint,superscriptaddress,twocolumn,preprintnumbers,floatfix,nofootinbib]{revtex4-1}

\usepackage[dvipdfm]{graphicx}
\usepackage{mathtools}
\usepackage{epstopdf}
\usepackage{multirow}
\usepackage{verbatim}
\usepackage{color}
\usepackage[dvipsnames]{xcolor}
\usepackage{slashed}
\usepackage{bbold}
\usepackage{fullpage}
\usepackage{float}
\usepackage[normalem]{ulem}
\usepackage[colorlinks, citecolor=blue, linkcolor=black]{hyperref}
\usepackage{amsfonts}
\usepackage{amsmath}
\usepackage{amssymb}
\hypersetup{pdfstartview=FitV,colorlinks=true,linkcolor=blue,citecolor=red,filecolor=black,urlcolor=blue}

\def\({\left(}
\def\){\right)}
\def\[{\left[}
\def\]{\right]}

\usepackage[parfill]{parskip}
\begin{document}
\sloppy


\vspace*{1mm}

\title{Measuring Lepton Flavor Violation at LHC}

\author{K. Asai}
\affiliation{Ochanomizu University, Otsuka, Bunkyo-ku, Tokyo, Japan}
\author{J. Fujimoto}
\affiliation{KEK, High Energy Accelerator Research Organization, Tsukuba, Japan}
\author{K. Kaneta}
\affiliation{Department of Mathematics, Tokyo Woman’s Christian University, Tokyo 167-8585, Japan
}
\author{Y. Kurihara}
\affiliation{KEK, High Energy Accelerator Research Organization, Tsukuba, Japan}
\author{S. Tsuno}
\affiliation{KEK, High Energy Accelerator Research Organization, Tsukuba, Japan}

\begin{abstract}
A new process with the lepton flavor violation (LFV) is presented in the setup for LHC. The LFV is induced by the one-loop effect through Higgs bosons in the framework of the type-III two Higgs doublet model. It is demonstrated that the vast of parameter space in the Yukawa sector could be accessed by current and future LHC experiments.
\end{abstract}

\maketitle

\setcounter{equation}{0}

\section{I. Introduction}

Lepton Flavor Violation (LFV) is not only a consequence of the nonzero neutrino masses and oscillations but also a tool to search for various types of theoretical models beyond the Standard Model (SM).
For instance, the SM prediction for $\mu\to e\gamma$ is too small to be observed in the foreseeable experiments \cite{Petcov:1976ff,Cheng:1980tp}.
Therefore, any signal of LFV gives some hint on new physics beyond the SM.
Indeed, it is well known that supersymmetric (SUSY) models generically give rise to LFV effects through soft SUSY breaking effects in the slepton sector \cite{borzumati1986,leontaris1986}.
In conjunction with the nonzero masses of the neutrinos, right-handed neutrinos are highly motivated particle that may explain not only the neutrino masses via the seesaw mechanism \cite{minkowski1977,yanagida1979,gell-mann1979,mohapatra1980,schechter1980,schechter1982} but also the baryon asymmetry of the Universe \cite{fukugita1986}.
By putting these particles into the SUSY framework, LFV in the slepton sector may be induced by radiative corrections, even when underlying physics behind the SUSY breaking has nothing to do with LFV \cite{hisano1995,hisano1996,hisano1999,casas2001,ellis2002,Haba:2012ai}.

In testing such LFV models, $\mu\to e\gamma$ often gives the most stringent constraint.
The SUSY model with right-handed neutrinos can be a typical example to get a feeling of the constraint, where the LFV appears at one-loop with slepton and chargino/neutralino inside the loop, and thus the amplitude is proportional to the soft term $\tilde m^2_{Lij}$ with $i\neq j$ and $i,j=1,2,3$ denoting the lepton-sector generation.
By taking $\tilde m$ as a typical SUSY particle mass, the branching ratio may roughly be estimated as ${\rm Br}_{l_i\to l_j\gamma}\sim \alpha^3|\tilde m^2_{Lij}|^2/\tilde m^8 G_F^2$.
The LFV soft term is generated through a self-energy diagram of sleptons, where the right-handed neutrinos come inside the loop with a neutrino Yukawa coupling $y_\nu$.
By neglecting details of the loop such as logarithmic piece and contributions from different type of soft terms etc., $\tilde m^2_{Lij}\sim (16\pi^2)^{-1}\tilde m^2 (y_\nu^\dagger y_\nu)_{ij}$ can be obtained, yielding, for instance, ${\rm Br}_{\mu\to e\gamma}\sim 10^{-7}|(y_\nu^\dagger y_\nu)_{21}|^2(m_W/\tilde m)^4$ which should be compared with the current limit ${\rm Br}_{\mu\to e\gamma}<4.2\times10^{-13}$ \cite{MEG:2016leq}.
Therefore, the LFV measurements have been one of the powerful tools to look for physics beyond the SM.

From the current experimental searches for the LFV processes, the most stringent constraint has been given to the LFV effects involving gauge interactions, such as $\mu\to e\gamma$.
On the other hand, it could be that searches for the LFV involving Yukawa interactions give a complementary path to probe new physics.
Collider experiments provide such opportunity that the both types of LFV processes can be explored simultaneously.

The LFV processes have been searched through the rare decay of the SM particles of $Z$ and SM Higgs boson \cite{ATLAS:2020zlz,ATLAS:2019pmk,ATLAS:2019old,CMS:2021rsq} as well as exotic particles \cite{ATLAS:2018mrn,ATLAS:2017xqs} at LHC. Their limits so far are ${\rm Br}_{Z\to e\tau}<8.1\times10^{-6}$, ${\rm Br}_{Z\to \mu\tau}<9.5\times10^{-6}$ \cite{ATLAS:2020zlz} and ${\rm Br}_{h_{SM}\to e\tau}<2.2\times10^{-3}$ and ${\rm Br}_{h_{SM}\to \mu\tau}<1.5\times10^{-3}$ \cite{CMS:2021rsq}, respectively. It should be noted that those studies had been carried by searching or measuring the resonance particles of $Z$ or $h_{SM}$.

In this paper, LFV processes of $W^+ W^- \to l_{i} l_{j}$ at LHC is investigated,  based on the type-III two Higgs doublet model (THDM) which provides a generic parametrization of LFV couplings in the Yukawa sector. Within the framework, such LFV processes may arise at one loop level mediated by heavy neutral and charged Higgs bosons while the tree level contributions are largely suppressed at the hypothesis with large tan$\beta$ region in THDM. It will be shown that, although the cross section is loop-suppressed, it is still accessible at future LHC runs, especially, for the parameter spaces where the extra Higgs bosons are at multi-TeV scales, and thus, complementary parameter spaces in LFV couplings can be covered.

The paper is organized as follows. Our framework and parametrization of the type-III two-Higgs doublet model are explained in sec. II. The one loop calculation and their event generation at LHC condition are described in sec. III and the numerical results follows in sec. IV. Finally, the feasibility study to constrain the relevant parameters on the LFV couplings is given in sec.V, then sec. VI is devoted to the discussion and conclusion.

\section{II. Model}

Among various possible sources for LFV, the LFV couplings in the Higgs sector with two Higgs doublet fields are considered in the rest of the paper. In the absence of a flavor symmetry, Higgs-mediated flavor changing neutral current (FCNC) often becomes problematic, since it is not always the case where the Yukawa couplings and fermion mass matrices can be simultaneously diagonalized.The problematic FCNC can be avoided if there is a $Z_2$ symmetry under which, for instance, only one of the two Higgs fields and the up-type quarks are odd parity so that the Higgs only gives masses to the up-type quarks~\cite{Glashow:1976nt}. This model is called type-II two Higgs doublet model, and the minimal supersymmetric SM (MSSM) falls into this class (at tree-level).

However, such flavor symmetry is often not guaranteed against radiative corrections. Indeed, in the MSSM, the SUSY breaking does not respect the flavor symmetry in general. Consequently non-holomorphic Yukawa couplings may appear in the low energy theory which turns out to be the so-called the type-III two Higgs doublet model.

Aside from the detail of the origin of the non-holomorphic Yukawa couplings, the (lepton sector) low effective theory may be written as
\begin{eqnarray}
    -{\cal L}_{\rm lep} &\simeq& \bar l_{Ri}\left[
    y_{li}\delta_{ij} H_d^\dagger + (y_{li}\kappa_{Rij}+\kappa_{Lij}y_{lj}) H_u^\dagger
    \right]L_j \nonumber\\
    &&+ h.c.,
\end{eqnarray}
where $i,j=e,\mu,\tau$, $H_u=(H_u^+~H_u^0)^T, H_d=(H_d^0~H_d^-)^T$, $\kappa_L$ and $\kappa_R$ parametrize the flavor off-diagonal contributions in the mass eigenstate basis of leptons.
After taking the mass eigenstates for the Higgs fields, the effective Yukawa interactions become  \cite{Babu:2002et,Brignole:2004ah,Kanemura:2005hr,Raidal:2008jk}
\begin{widetext}
\begin{eqnarray}
	-{\cal L}_{\rm LFC} &=& (2G_F^2)^{1/4} m_{li}
	\left[
	(\overline l_{Ri} l_{Li})\left(-\frac{s_\alpha}{c_\beta}h^0+\frac{c_\alpha}{c_\beta}H^0+it_\beta A^0\right)-\sqrt{2}t_\beta (\overline l_{Ri}\nu_{Li})H^-+h.c.
	\right],
	\label{eq:LFC}\\
	-{\cal L}_{\rm LFV} &=& (2G_F^2)^{1/4} \frac{m_{li}}{c_\beta^2}\left[
	\kappa_{Rij}\overline l_{Li}l_{Rj}(c_{\beta-\alpha}h^0-s_{\beta-\alpha}H^0-iA^0)
	+\sqrt{2}\kappa_{Rij}\overline l_{Rj}\nu_{Li}H^-+h.c.
	+(R\leftrightarrow L)
	\right],
	\label{eq:LFV}
\end{eqnarray}
\end{widetext}
where $s_x\equiv \sin x$, $c_x\equiv\cos x$, $t_x\equiv \tan x$, ${\cal L}_{\rm LFC}$ and ${\cal L}_{\rm LFV}$ denote the (charged) lepton flavor conserved and violated pieces of the Lagrangian, respectively.
Note that the charged leptons are taken as the mass eigenstate, while the neutrinos are in the interaction basis, and thus a unitary matrix $U^{PMNS}$ (Pontecorvo-Maki-Nakagawa-Sakata matrix) appears when taking their mass basis. 

In the following analysis, $\kappa_L=0$ and $\kappa_R\equiv\kappa$ are assumed for simplicity. Table.\ref{tab:ffhmssm} summarizes the coupling constants for Higgs bosons coupled with fermions with LFV ($i$ $\neq$ $j$) as well as LF-conserved couplings ($i$ $=$ $j$), where indices $i$ and $j$ indicate a generation of the leptons. 
For the other relevant Higgs couplings, the notation follows Ref. \cite{Kuroda:1999ks}.

\begin{table}[tbhp]
\begin{center}
\begin{tabular}{l|cc} \hline \hline
  Vertex        & LF-conserved & non LF-conserved \\
               & ($i=j$) & ($i\neq j$) \\ \hline
  \multirow{4}{*}{\includegraphics[height=1.2cm]{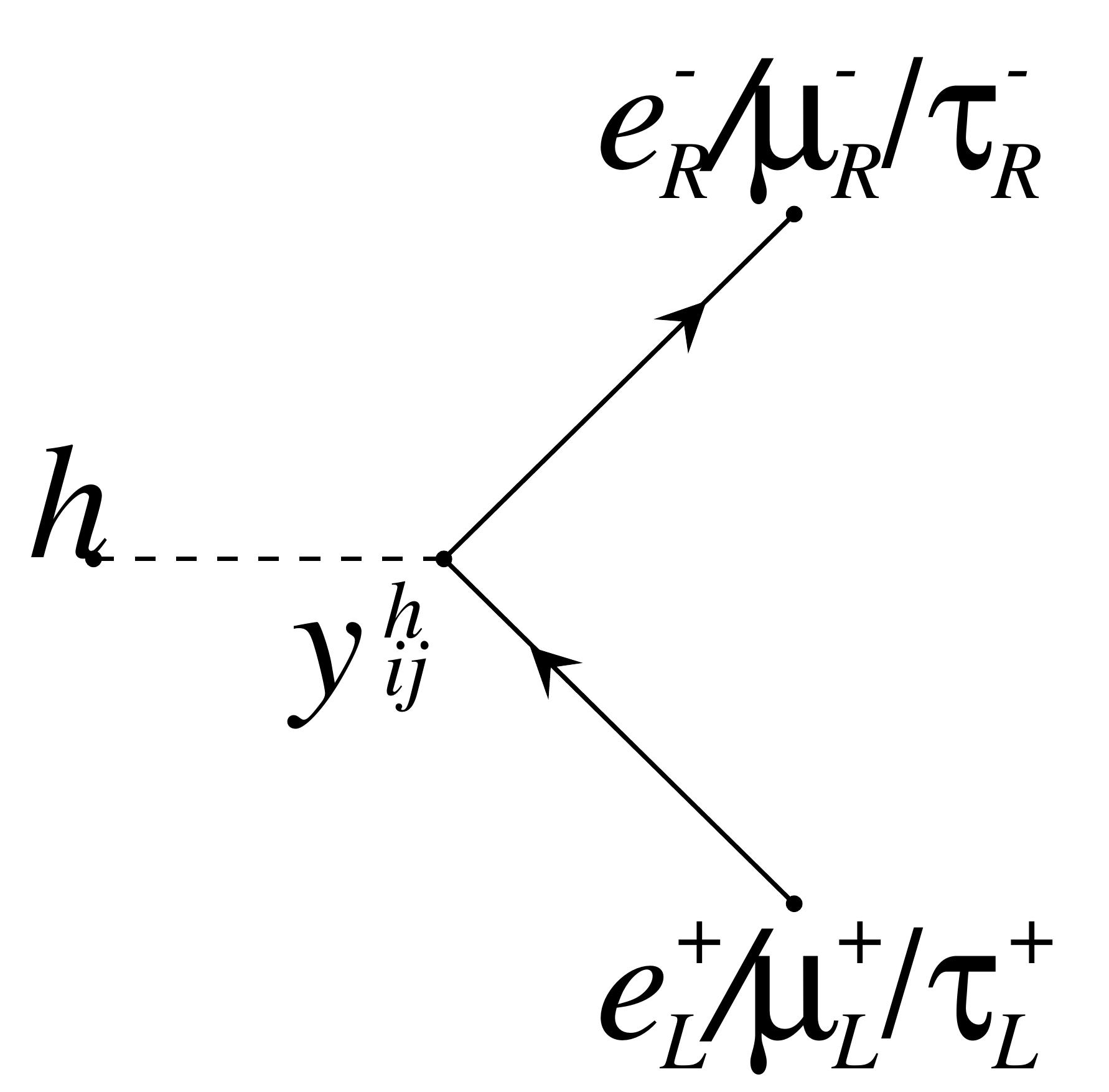}} & & \\
  & ${ \begin{aligned}[t]
    i m_{\ell_{R}} \left( \frac{s_{\alpha}}{c_{\beta}} \right)
  \end{aligned} }$
  & ${ \begin{aligned}[t]
    i m_{\ell_{R}} \left( \frac{\kappa_{ij} c_{\beta-\alpha}}{c^{2}_{\beta}} \right)
  \end{aligned} }$ \\
  & & \\ \hline

  \multirow{4}{*}{\includegraphics[height=1.2cm]{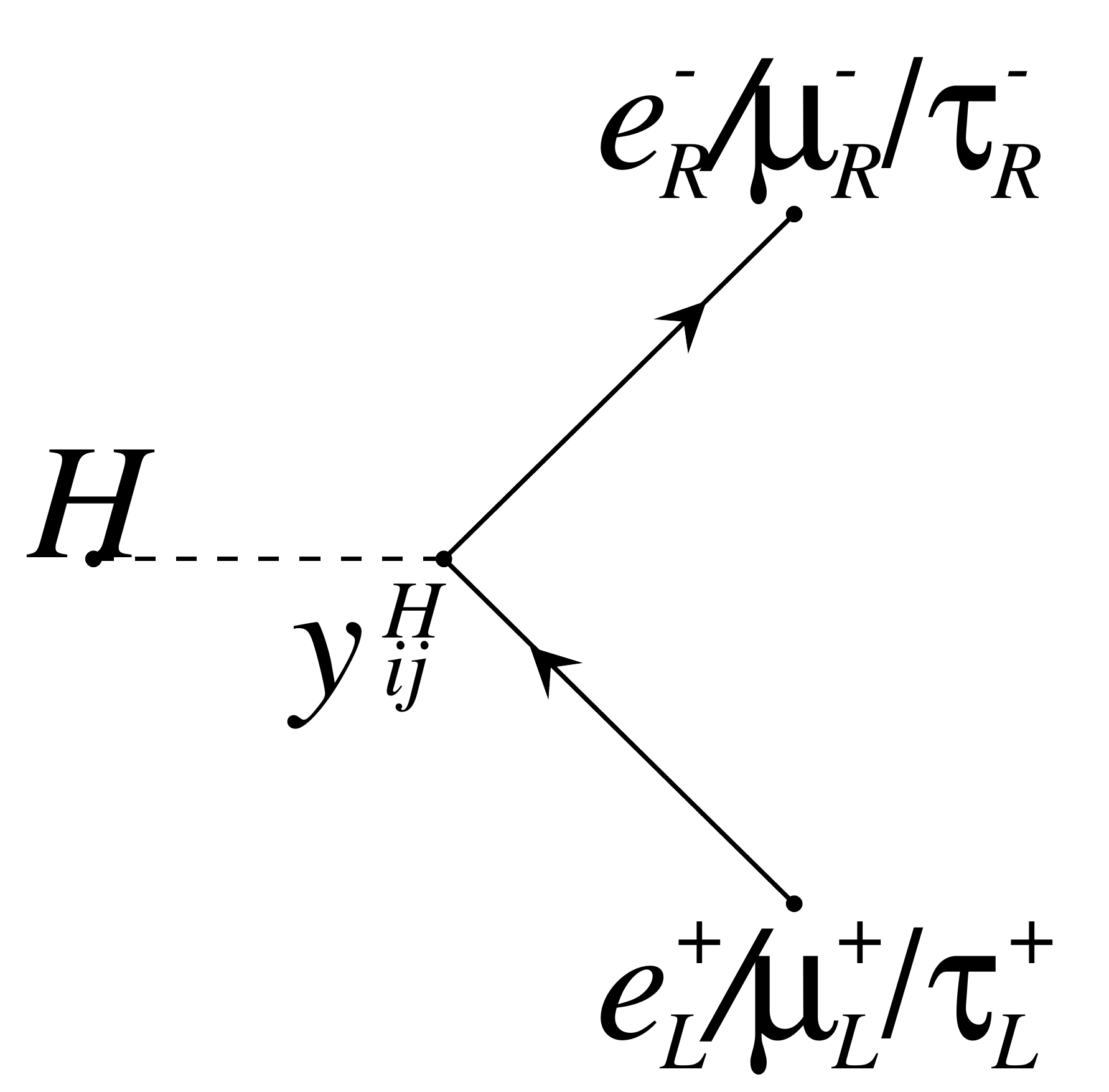}} & & \\
  & ${ \begin{aligned}[t]
    -i m_{\ell_{R}} \left( \frac{c_{\alpha}}{c_{\beta}} \right)
  \end{aligned} }$
  & ${ \begin{aligned}[t]
    -i m_{\ell_{R}} \left( \frac{\kappa_{ij} s_{\beta-\alpha}}{c^{2}_{\beta}} \right)
  \end{aligned} }$ \\
  & & \\ \hline

  \multirow{4}{*}{\includegraphics[height=1.2cm]{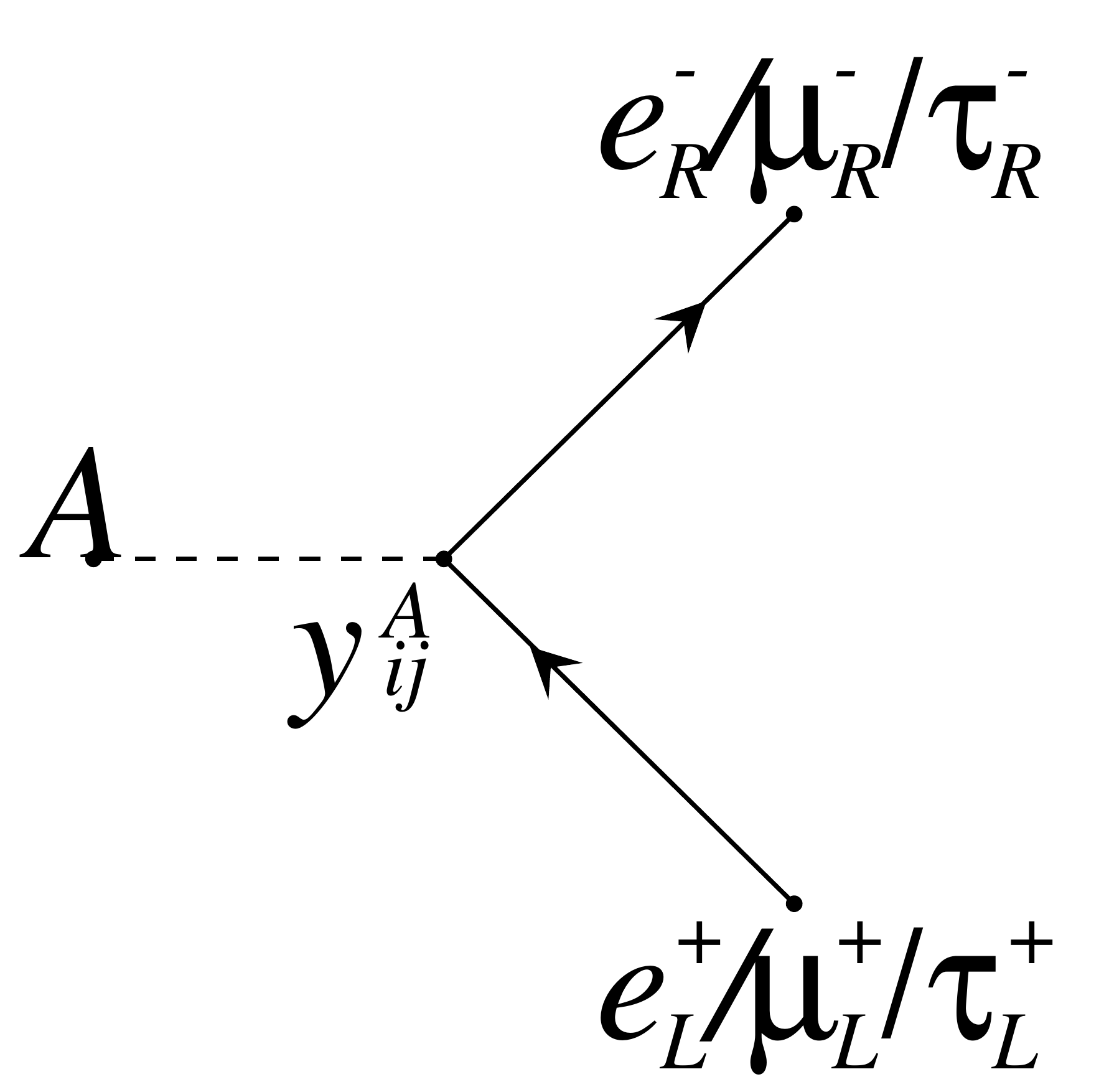}} & & \\
  & ${ \begin{aligned}[t]
    m_{\ell_{R}} t_{\beta}
  \end{aligned} }$
  & ${ \begin{aligned}[t]
    m_{\ell_{R}} \left( \frac{\kappa_{ij}}{c^{2}_{\beta}} \right)
  \end{aligned} }$ \\
  & & \\ \hline

  \multirow{4}{*}{\includegraphics[height=1.2cm]{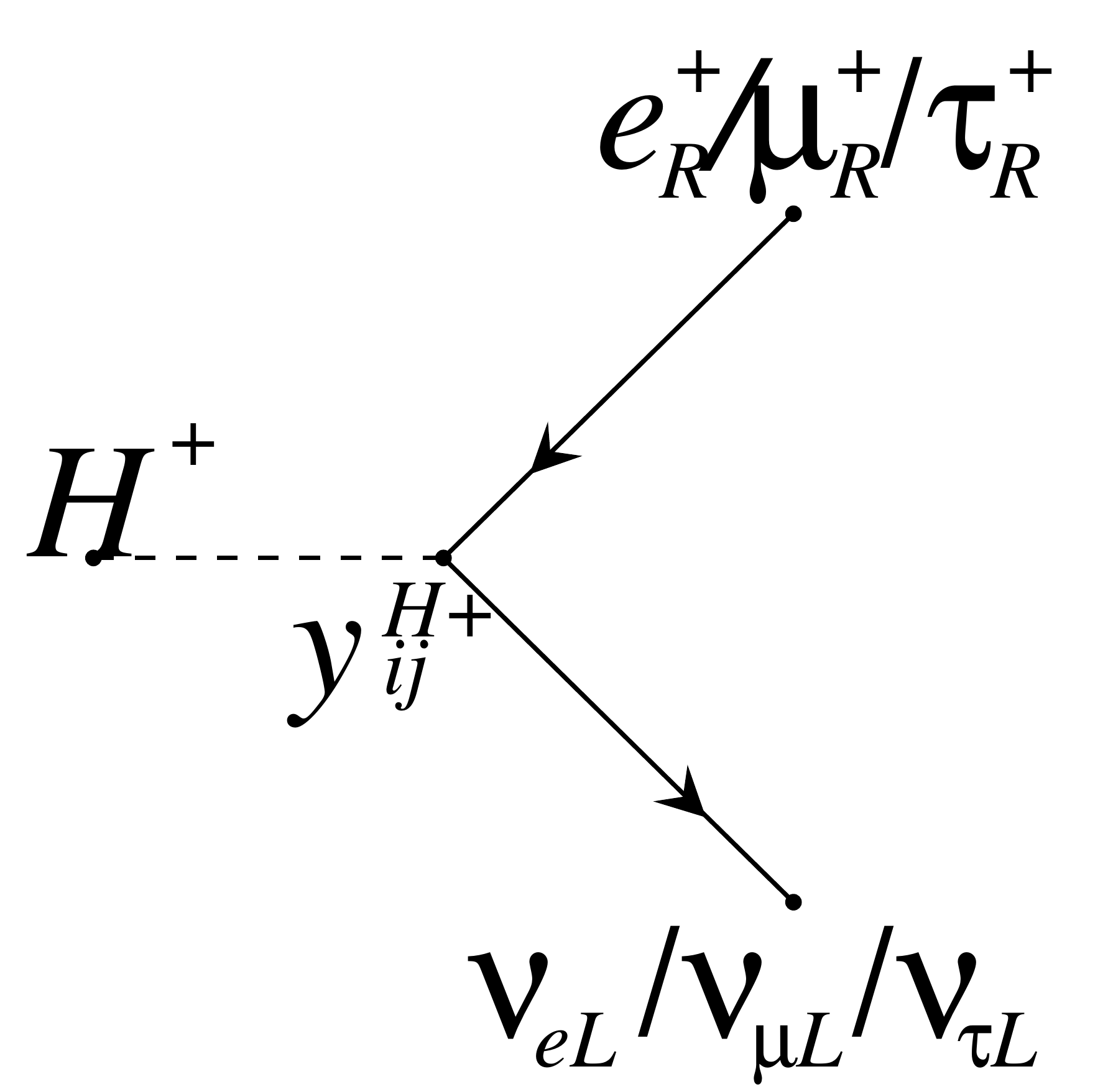}} & & \\
  & ${ \begin{aligned}[t]
    i |U^{\mathrm{PMNS}}_{ij}| m_{\ell_{R}} t_{\beta}
  \end{aligned} }$
  & ${ \begin{aligned}[t]
    i |U^{\mathrm{PMNS}}_{ij}| m_{\ell_{R}} \left( \frac{\kappa_{ij}}{c^{2}_{\beta}} \right)
  \end{aligned} }$ \\
  & & \\ \hline\hline
\end{tabular}
\caption{
  Coupling constants of h, H, A and $H^{+}$ bosons with leptons (index $i (j) = 1,2,3$), where $(2G_{F}^{2})^{\frac{1}{4}}$ is omitted.
\label{tab:ffhmssm}
}
\end{center}
\end{table}

\section{III. Set up}

At the decoupling limit, where the lighter Higgs boson $h$ is close to the SM Higgs boson ($c_{\alpha} \approx 0$, $m_{H/A} > 500 \mathrm{GeV}$, the heavier Higgs bosons $H$ and $A$ have a sizeable couplings of $H \to l_{i} l_{j} (i \neq j)$ proportional to the $\kappa_{ij}/c^{2}_{\beta}$ when the $t_{\beta}$ is large. However, the coupling with gauge bosons ($W^+ W^- \to H$) is largely suppressed. 
The $A$ boson coupling with gauge bosons even does not exist. Thus, the $s$-channel mode at tree level is largely suppressed. The higher order diagrams instead play an important role in the LFV.

To evaluate such a higher order LFV interactions, the effective 1-loop vertices are constructed. The calculation is made by the helicity amplitude method based on CHANEL \cite{Tanaka:1989gu} library. First, the tree-level amplitude is constructed by GRACE system \cite{Yuasa:1999rg}, which is an automatic code generation program for given initial and final state particles. This provides all possible Feynman diagrams with gauge invariant set and allows to calculate the squared amplitude for those diagrams. The LFV interactions are not introduced here since the system only assumes the SM interactions in the model. After the code generation, the tree level vertices are replaced with the corresponding LFV effective 1-loop vertices. Therefore, the base process at the starting point to be produced by the GRACE system is the 2 $\rightarrow$ 4 body process,
\begin{equation}
  q_{1} \; + \; q_{2} \; \rightarrow \; \mu^{+} \; + \; \mu^{-} \; + \; q_{3} \; + \; q_{4} \; ,
\end{equation}
where $q_{n}~(n=1,...,5)$ is a quark flavor except top-quark that allows possible combination for hadron-hadron collisions. At this level, about 20,000 diagrams are generated under unitary gauge. Then GR@PPA package \cite{Tsuno:2006cu}, an extension of the GRACE system for hadron colliders, applies the diagram reduction taking into account for the charge conjugate, unification of flavor-blind interaction and parity-conservation for the exchange of the initial colliding partons, and also connects to the parton density function for colliding hadrons (PDF \cite{Whalley:2005nh}). This package finally provides about 100 core diagrams to be calculated.

The next step is to replace with the effective vertices. The most general structure of a vertex formula with vector current is given as
\begin{equation}
  \Gamma_{\mu} \; = \; (A + B \gamma_{5}) \gamma_{\mu} + (C + D \gamma_{5})(p' \pm p)_{\mu} \; ,
  \label{formula_coef}
\end{equation}
where $p$ and $p'$ are momenta of the external fermions and coefficients $A$ to $D$ are given by the loop integration functions. The first term of $(A + B \gamma_{5})$ corresponds to the tree level vector current vertex proportional to $\gamma_{\mu}$ and second term is a scalar vertex coupled with fermions. The loop correction is in general decomposed by the vector and scalar interactions.

\begin{figure}[hptb]
\begin{center}
\includegraphics[height=4cm]{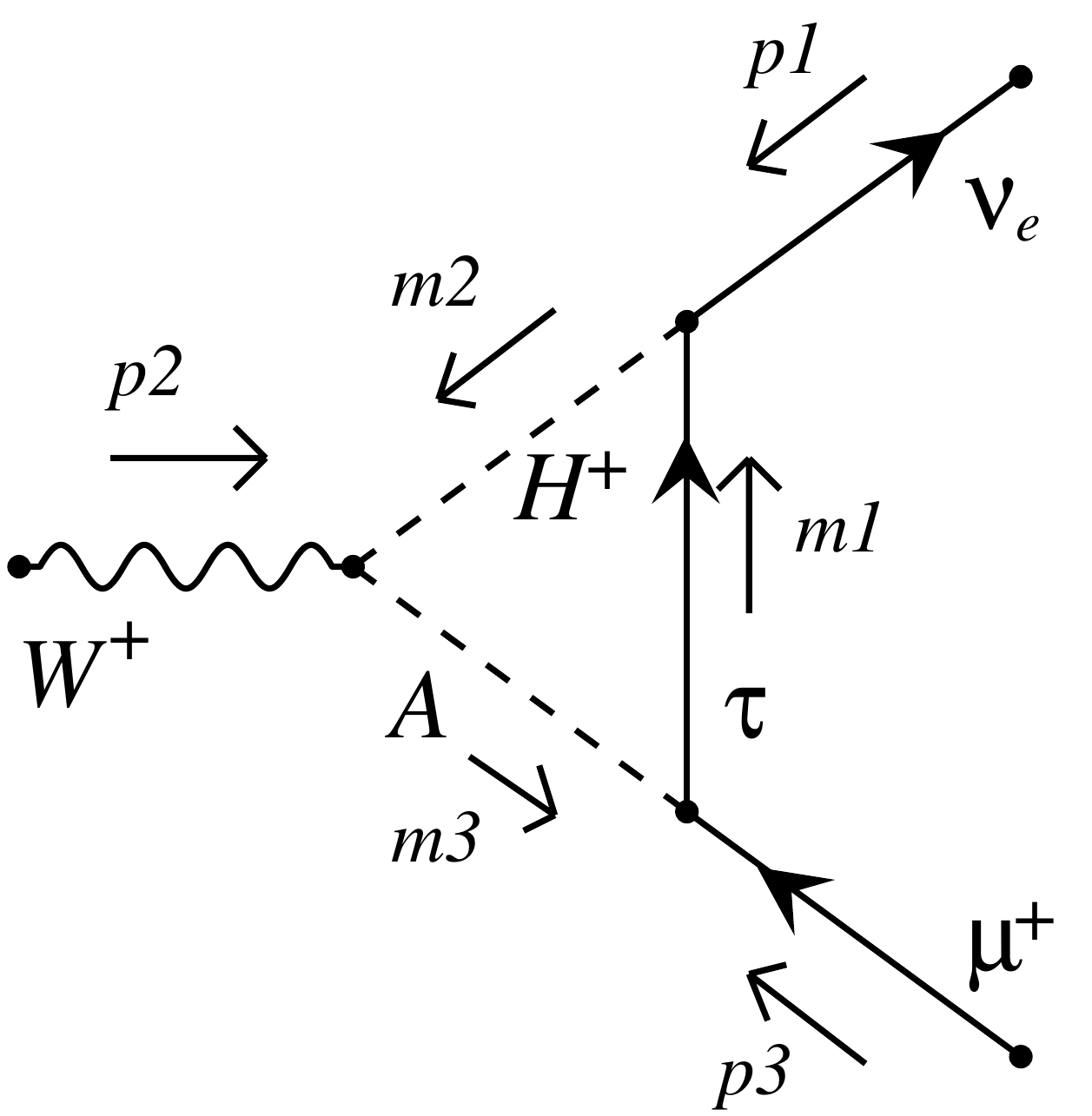}
\caption{
Triangle loop diagram with two scalar bosons ($A$ and $H^{+}$) and one fermion ($\tau$) in the loop on one external vector boson ($W$) and two external fermions ($\mu^{+}$ and $\nu_{e}$).
}
\label{triangle_w2hpaf.v2}
\end{center}
\end{figure}

Considering Fig.\ref{triangle_w2hpaf.v2} as an example of the effective vertex with LFV interaction, the 1-loop amplitude is expressed as
\begin{equation}
  \begin{aligned}
    \Gamma(W\rightarrow\mu\nu_{e})_{\mu} = \qquad \qquad \qquad \qquad \qquad \qquad \qquad \\
   \;  \\
    = \frac{1}{16\pi^{2}} \bigg[ \left( \int_{0}^{1} \int_{0}^{1} y \mathrm{ln}D(x,y) dx dy - \frac{1}{2} C_{UV} \right) \gamma_{\mu} \\
    \;  \\
    + \int_{0}^{1} \int_{0}^{1} \frac{y dx dy}{D(x,y)} ( m_{1} - \bar{p}\hspace{-1.8mm}/\hspace{0.3mm} ) ( -2 \bar{p} + 2 p_{1} + p_{2} )_{\mu} \bigg] ,
  \end{aligned}
  \label{formula_w2hpaf}
\end{equation}
where the coupling constants are omitted and $C_{UV}$ ($\equiv \frac{1}{\varepsilon}+\mathrm{ln}4\pi$) is an ultraviolet divergent part. The $D(x,y)$ is the outcome of the Feynman integral defined as
\begin{equation}
  D(x,y) = -y \bar{p} \{ (1-x) p_{1} + p_{2} \} + xy p_{1} p_{2} + \bar{m}^{2} ,
\end{equation}
with
\begin{equation}
\begin{aligned}
  \bar{p} = ( 1 + xy - y ) p_{1} + ( 1 - y ) p_{2} \qquad \\
  \bar{m}^{2} = y \{ ( 1 - x ) m_{1}^{2} + x m_{2}^{2} \} + ( 1 - y ) m_{3}^{2} .
  \end{aligned}
  \end{equation}
The numerical integration is performed inside code. The output is checked with LoopTools \cite{HAHN1999153} and our previous study \cite{Aoki:1982ed}. All relevant vertex formula's and coupling constants with the LFV interactions are implemented with same manner. The relevant tree-level vertex with $W$ $\rightarrow$ $\mu\nu_{\mu}$ is now replaced with Eq.(\ref{formula_w2hpaf})
together with the corresponding coupling constants. The typical order of such loop correction is
\begin{equation}
   \Gamma_{\mu} \; \sim \; 10^{-2} \gamma_{\mu} + 10^{-4} p_{\mu} \; ,
\end{equation}
for $m_{A}$ $=$ 1 TeV, $t_{\beta}$ $=$ 40, $\kappa_{23}$ $=$ $\kappa_{13}$ $=$ 0.1 at LHC condition. Each coefficient corresponds to the parameters $A$($=B$) and $C$($=D$) in Eq.(\ref{formula_coef}). Those parameters varies to the input momenta used in the vertex calculation. The outgoing leptons ($\mu^{+}\mu^{-}$) are also replaced with the relevant lepton flavors, that results in the LFV in the end.

\begin{figure}[hptb]
\begin{center}
\includegraphics[width=5.0cm]{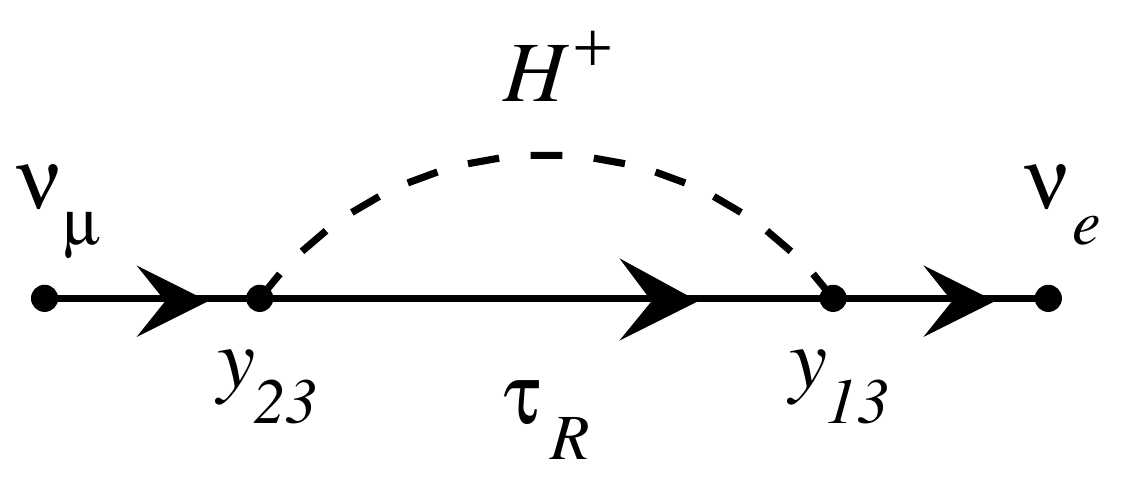}
\vspace*{0.5cm}
\caption{
Self-energy diagram that the muon neutrino ($\nu_{\mu}$) changes the flavor to electron neutrino ($\nu_{e}$) through charged Higgs boson ($H^{+}$) in the loop. Note that the Yukawa interaction couples with right-handed tau lepton ($\tau_{R}$)
}
\label{self_t_st_m.v2}
\end{center}
\end{figure}

Another type of the LFV process is through the self-energy diagrams. Typical diagram is shown in Fig.\ref{self_t_st_m.v2}.
This diagram is known to have a logarithm mass-dependence ($\sim$log($m_{H}$)) in the loop structure. Therefore, the amplitude diverges as an increase of the input Higgs boson mass. To avoid such divergence, the renormalization scale $\mu_{R}$ is set to be $m_{A}$ ($\sim m_{H+}$ at $m_{A}$ $>$ 500 GeV) to cancel the mass-dependence. This is interpreted that the purturbation is only valid at this scale. This choice minimizes the contributions from the flavor-changing self-energy diagram. Thus, predicts minimal production cross sections.

Soft-photon in the loop is a source of a logarithmic divergence and could be canceled by the real photon emission process at tree level. But such diagrams are raised by the $s$-channel process, where $h$ or $H$ bosons are propagated. Since those diagrams have either of the coupling of the $h \to ll'$ or $WW \to H$, those contributions are negligibly small. Therefore, the soft-photon term is neglected in the calculation. For the same reason, the box-type diagrams are also ignored.

\begin{figure}[htbp]
\begin{center}
\includegraphics[width=6cm]{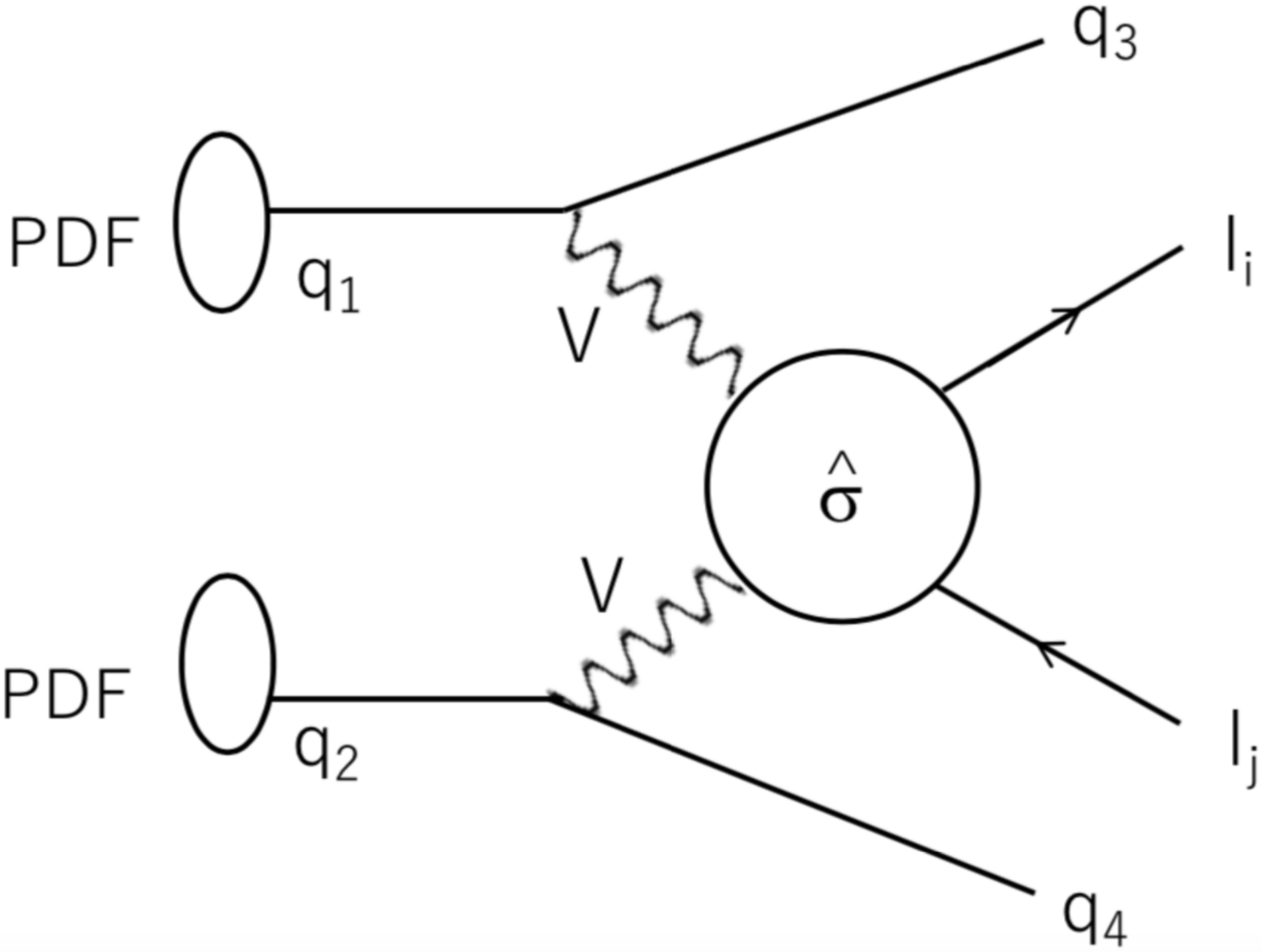}
\caption{
Schematic picture for the diagram calculation. The one-loop effect is taken into account in the $\hat{\sigma}$, while the initial and final state quarks are based on the tree level diagram calculation.
}
\label{fig:lfvprodview}
\end{center}
\end{figure}

Though the $W^+ W^- \to l_{i} l_{j}$ (and $Z/\gamma Z/\gamma \to l_{i} l_{j}$) is produced through the vector boson scattering process at LHC, the loop corrections are applied to the vertices in $VV$ $\rightarrow$ $\mu^{+}\mu^{-}$ process. At decoupling limit in THDM ($m_{A}$ $>$ 500 GeV, $t_{\beta}$ $>$ 10), the tree level Higgs decay mode into the LFV is suppressed and found to be less than 1\% contribution to the 1-loop diagram calculation, and thus, the calculation is performed with the 1-loop order only. The schematic view is illustrated in the Fig.\ref{fig:lfvprodview}. The Matrix Element is based on the 2 $\rightarrow$ 4 body process and the core part of the $VV$ $\rightarrow$ $l_{i}^{+}l_{j}^{-}$ interactions is based on 1-loop order calculation.

The production cross section with the lepton $l_{i}$, $l_{j}$ ($i\neq j$) is thus expressed as 
\begin{eqnarray}
    \lefteqn{ \sigma(pp\rightarrow l_{i}l_{j}+qq+X) } \\ \nonumber
    & = & \int x_{1}x_{2}f_{1}(x_{1})f_{2}(x_{2})\hat{\sigma}(q^{2})dx_{1}dx_{2}d\Psi ,
\end{eqnarray}
where $x_{1}$ and $x_{2}$ are the momentum fraction of the PDF $f_{1}$ and $f_{2}$ respectively. All combination of the incoming and outgoing quarks is taken into account in the calculation. The BASES/SPRING package \cite{Kawabata:1995th} handles numerical integration for the full-phase space mapping and the unweighted event. The 4-vector information for the initial and final state particles are stored with common format in the file \cite{ALWALL2007300}. Such file is interfaced by hadronization packages in later stage to simulate realistic events at LHC.

\section{IV. Result}

\begin{figure}[htbp]
\begin{center}
\includegraphics[width=7.5cm]{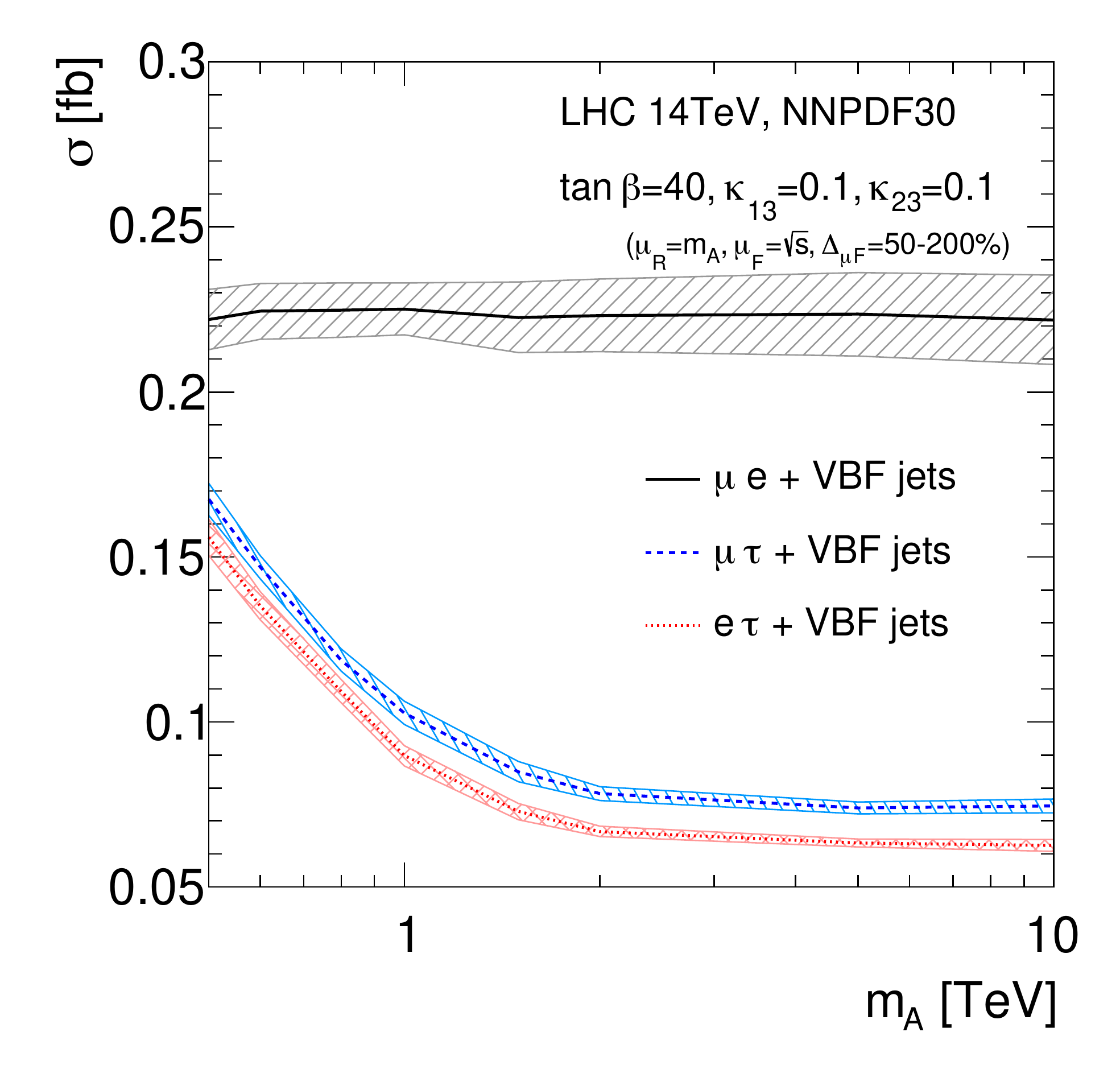}
\caption{
Production cross section at LHC 14 TeV condition at $t_{\beta}$ $=$ 40 as a function of $m_{A}$ ($\approx$ $m_{H}$ ($m_{H+}$)) for each LFV mode, $\mu e$, $\mu \tau$ and $e \tau$, respectively. The $\kappa$ values denoted in the figure. The systematic band is expressed as a variation of the different factorization scale of 50\% to 200\%.
}
\label{fig:xsec_massdep}
\end{center}
\end{figure}

The production cross section is presented as a function of $m_{A}$ ($\approx$ $m_{H}$ ($m_{H+}$)) at $t_{\beta}=40$ and $\kappa_{13}=\kappa_{23}=0.1$ with LHC 14 TeV condition in Fig.\ref{fig:xsec_massdep} for each LFV mode, $\mu e$, $\mu \tau$ and $e \tau$, respectively. In the calculation, $\overline{\rm MS}$ scheme is used. The renormalization scale is fixed at $\mu_{R} = m_{A}$, while the factorization scale $\mu_{F}$ is set as (square-root of) the invariant mass of the incoming partons ($\sqrt{\hat{s}}$) with 50\% to 200\% systematic variation as uncertainty, where PDF set ({\tt NNPDF30\_lo\_as\_0118}) is used \cite{Whalley:2005nh}. The following physics parameters are also used,
\begin{equation}
  \begin{aligned}
  \mathrm{EW} \; & \mathrm{parameters}, \\
 \; & m_{W} = 80.419 \; \mathrm{GeV}, m_{Z} = 91.188 \; \mathrm{GeV}, \\
    & m_{h} = 125.0 \; \mathrm{GeV}, \; \alpha_{em} = 1/128.07,
\end{aligned} \nonumber
\label{physparam}
\end{equation}
and for neutrini mixing parameters,    
\begin{equation}
  \begin{aligned}
  \mathrm{Normal} \; & \mathrm{ordering}, \\
\theta_{12}=33.44^{o}, & \; \theta_{13}=8.57^{o}, \; \theta_{23}=49.20^{o}, \delta_{CP} = 197^{o}, \\
  \mathrm{Inverted} \; & \mathrm{ordering}, \\
  \theta_{12}=33.45^{o}, & \; \theta_{13}=8.60^{o}, \; \theta_{23}=49.30^{o}, \delta_{CP}=282^{o},
  \end{aligned} \nonumber
\label{physparam2}
\end{equation}
where $m_{W}$, $m_{Z}$ and $m_{h}$ are masses of $W$, $Z$, and the SM Higgs bosons, respectively. The $\alpha_{em}$ is a fine structure constant defined at $m_{Z}$. The $\theta_{12}$, $\theta_{13}$, $\theta_{23}$ and $\delta_{CP}$ are the neutrino mixing parameters with normal (inverted) ordering taken from the latest combined results \cite{Esteban:2020cvm}. The following kinematical cuts are applied in the calculation,
\begin{equation}
  \begin{aligned}
  \mathrm{for} \; & \mathrm{leptons}, \\
& p_{T} > 15 \: \mathrm{GeV}, \; |\eta| < 2.5, m_{ll} > 200 \: \mathrm{GeV}, \\
  \mathrm{for} \; & \mathrm{outgoing} \; \mathrm{quarks}, \\
 & p_{T} > 20 \: \mathrm{GeV}, \; |\eta| < 4.5, m_{jj} > 300 \: \mathrm{GeV},
  \end{aligned} \nonumber
  \label{gencuts}
\end{equation}
where any leptons and jets should be separated by $\Delta R_{ll(lj)}$ $>$ 0.2 and jets must be separated by $\Delta R_{jj}$ $>$ 0.4.

\begin{figure}[htbp]
\begin{center}
\begin{tabular}{cc}
\includegraphics[width=3.5cm]{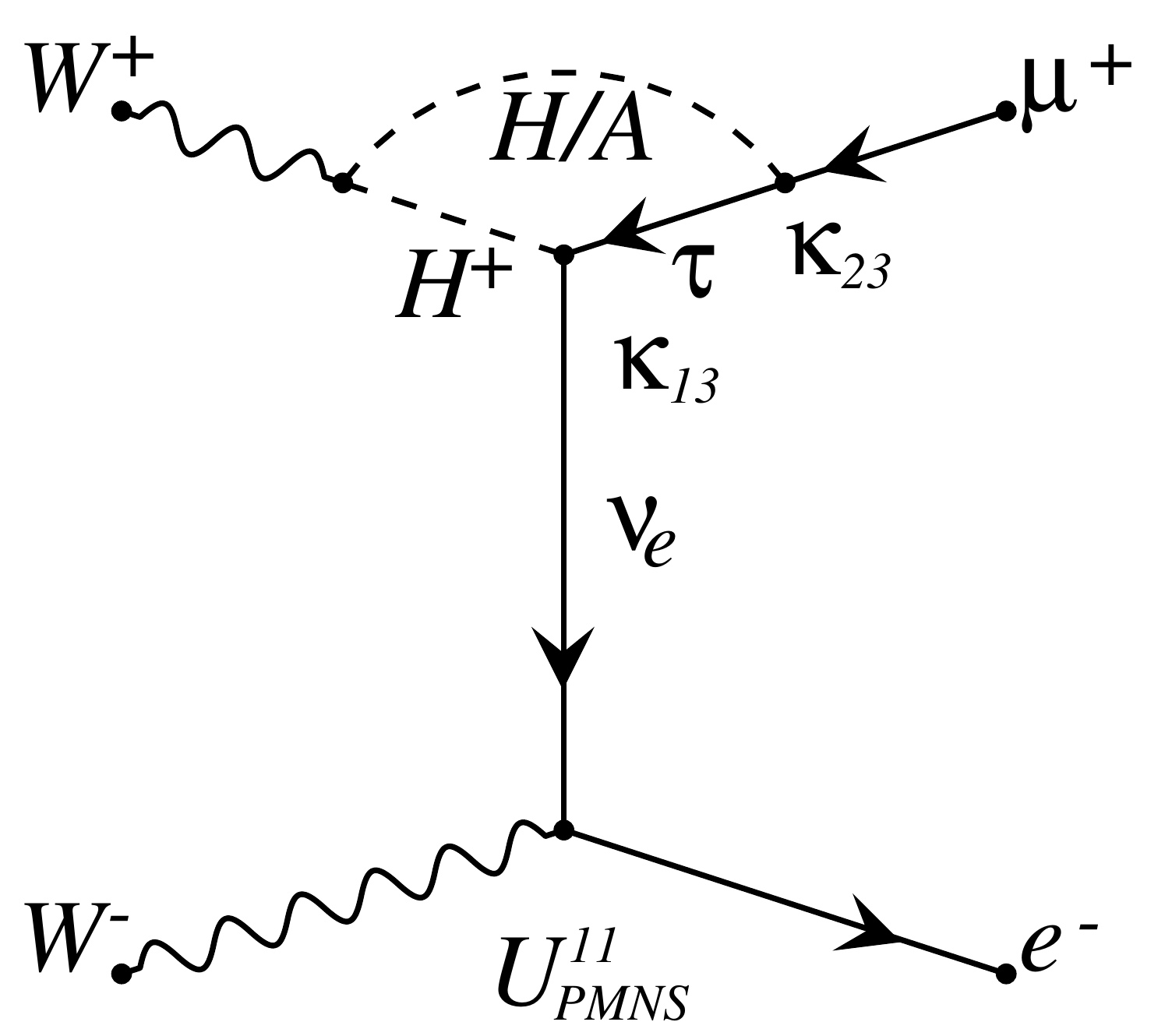} &
\includegraphics[width=3.5cm]{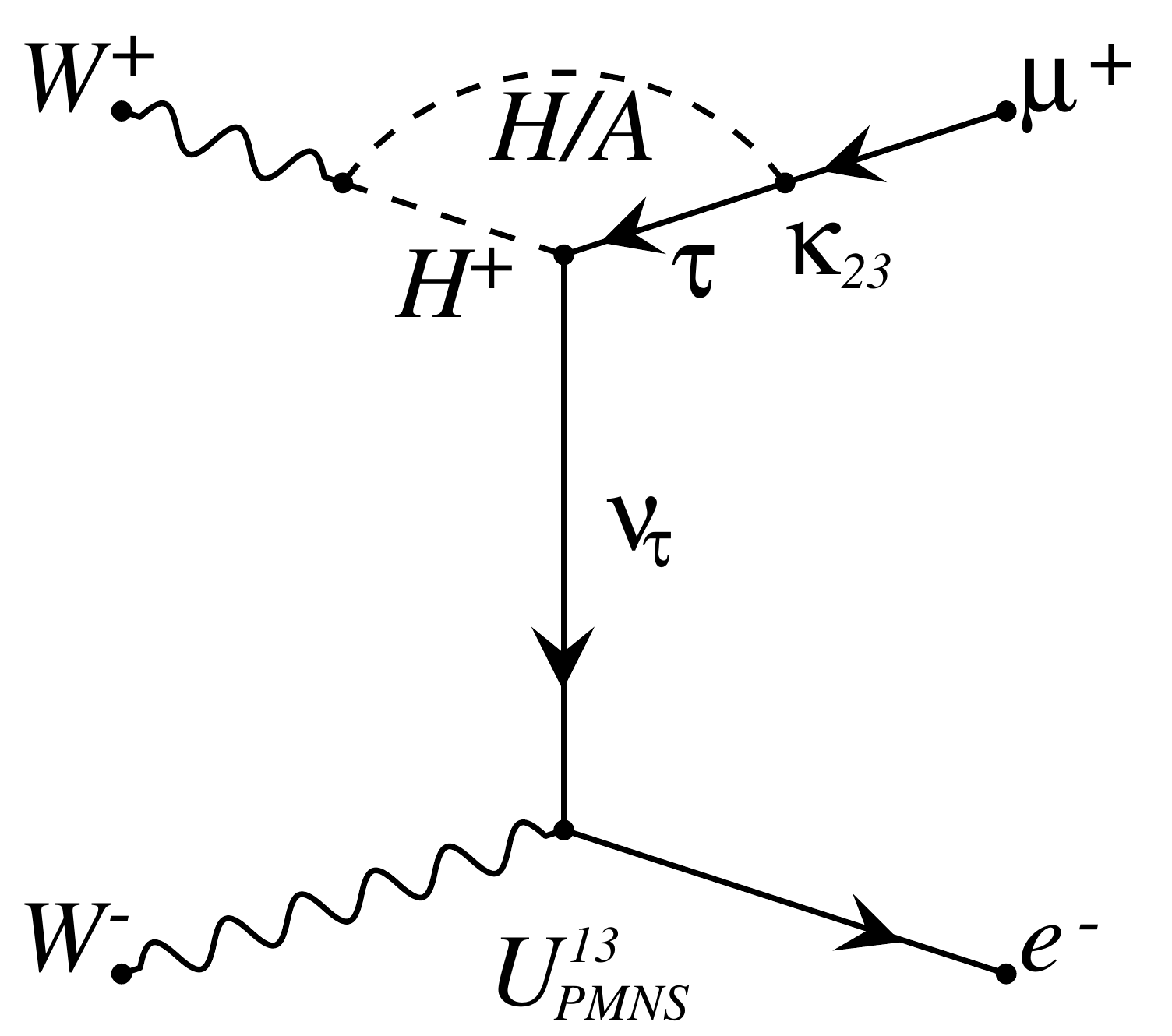} \\
(a) & (b) \\
\includegraphics[width=3.5cm]{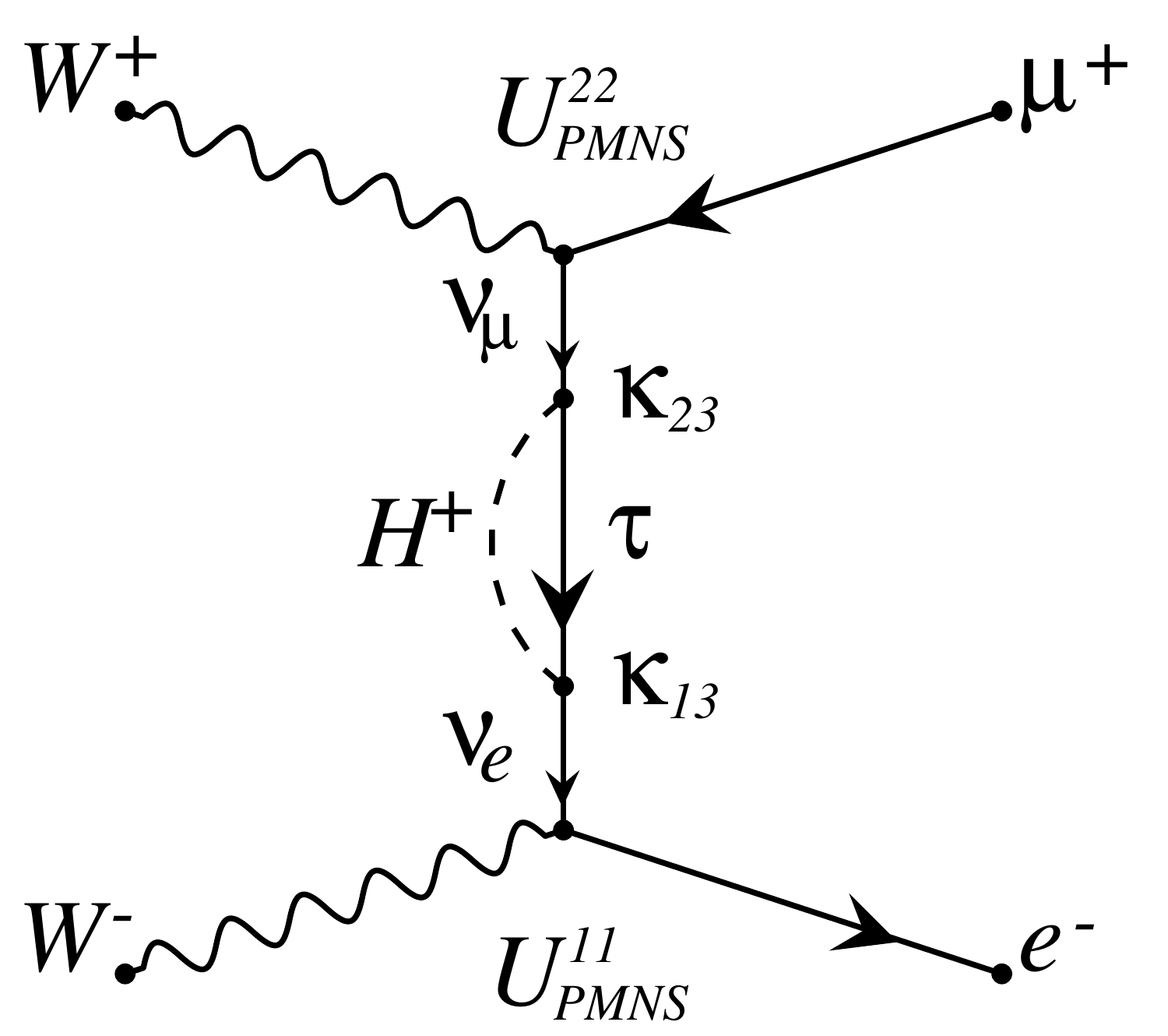} &
\includegraphics[width=3.5cm]{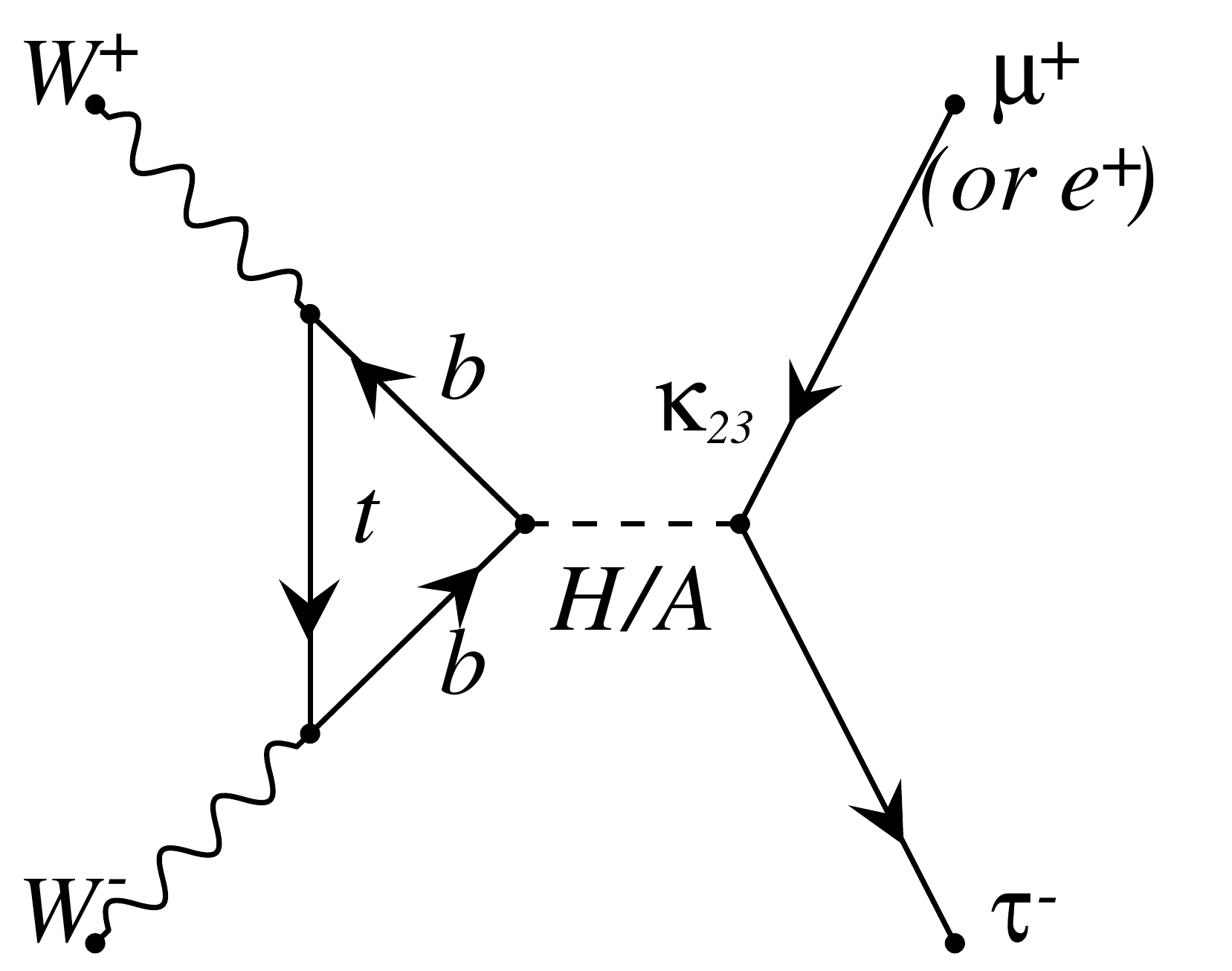} \\
 (c) & (d) \\
\end{tabular}
\caption{
Leading diagrams contributed in the production cross section. The flavor exchange occurs at the triangle loop vertex through charged Higgs boson (a) while it happens at the tree level vertex in the $W$ boson coupled with leptons through PMNS mixing matrix (b). The self-energy diagram in the $t$-channel neutrino mixing is not negligible by a given sizeable $\kappa$ and $\mathrm{tan}\beta$ parameters (c). The $s$-channel diagram with a fermion loop (d) only exist in the $\mu\tau$ and $e\tau$ final states but not in $\mu e$ final state because of negligible Yukawa couplings with $e$ or $\mu$.
}
\label{fig:leading_diagrams}
\end{center}
\end{figure}

The cross sections are stable at high $m_{A}$ region due to the fixed renormalization scale of $\mu_{R}$ $=$ $m_{A}$. This cross section gives the lower limit that minimizes the contribution from the self-energy divergence according to the input Higgs masses. Ignoring the interference between diagrams, the leading diagrams in the production are extracted as presented in Fig.\ref{fig:leading_diagrams}. In general, any combinations that have couplings with $h_{SM}$ $\rightarrow$ LFV or $WW$ $\rightarrow$ $H/A$ are largely suppressed by the decoupling condition. Thus, the $s$-channel diagrams do not contribute. This is why the tree-level direct production process in the neutral (non-LFV) MSSM Higgs boson searches do not have the VBF contribution. Meanwhile, the $t$-channel diagrams are dominant in LFV process through the loop contribution.

The flavor exchange occurs at the triangle loop vertex through charged Higgs boson (Fig.\ref{fig:leading_diagrams} (a)) while it happens at the tree level vertex in the $W$ boson coupled with leptons through PMNS mixing matrix (Fig.\ref{fig:leading_diagrams} (b)). The self-energy diagram in the $t$-channel neutrino mixing is not negligible according to an input $\kappa$ and $t_{\beta}$ parameters (Fig.\ref{fig:leading_diagrams} (c)). The neutrino mixing parameter plays an important role in the LFV. Since the flavor exchange at the tree level vertex in the $W$ boson account only at once in the diagram, the GIM suppression \footnote{Even number of flavor-exchanges by the $W$-boson undergoes GIM suppression by imposing the unitary condition of $\sum_{k=1,3}|U_{PMNS}^{ik}|^{*} \cdot |U_{PMNS}^{jk}|$ $=$ 0 where $i$ $\neq$ $j$. The off-diagonal elements are canceled out, thus no LFV occurs.} can not work in this case. The non-unitary structure in the mixing by the combination of $\kappa_{ij}$ and $U_{PMNS}^{ij}$ determines the sizeable contribution of the LFV in this process. At large $t_{\beta}$, the $\mu e$ final state is enhanced by the coupling structure by $\kappa_{13}\kappa_{23}/c^{4}_{\beta}$, while this relation is opposite at low $t_{\beta}$. Also, the cross sections decrease as $m_{A}$ increases for $\mu\tau$ and $e\tau$ final states while it is stable for the $\mu e$ final state due to lack of the $s$-channel contributions with a fermion loop in the $\mu e$ final state because the Yukawa coupling with $e$ or $\mu$ is negligible (Fig.\ref{fig:leading_diagrams} (d)). The $s$-channel contributions in $\mu\tau$ and $e\tau$ final states are visible up to $m_{A}$ $=$ 2 TeV at LHC condition. Table.\ref{tab:xsecsummary} summarizes the production cross sections with various parameter space for normal and inverted ordering of the neutrino mixing matrix.

The production cross sections depend on the $t_{\beta}$. The dependence is more pronounced in the $\mu e$ final state that contributes by a factor $1/c^{4}_{\beta}$ in the diagram (c). Then, the cross section becomes smaller than those in $\mu\tau$ and $e\tau$ at $t_{\beta}$ $\sim$ 15 since the enhancement by the $t_{\beta}$ is canceled by the $\kappa$ parameters ($\kappa t_{\beta}$ $\lesssim$ 1).

The $\kappa$ parameters are also scanned at the fixed $m_{A}$(=2TeV) and $t_{\beta}$(=40). Focusing on the diagrams (a) and (b), the asymmetric parameterization of the $\kappa_{13}$ and $\kappa_{23}$ gives rise to not only an asymmetric production rate between $\mu\tau$ and $e\tau$ final states, but also asymmetric contributions between diagrams. As summarized in Table \ref{tab:xsecsummary}, the $\kappa_{13}$($\kappa_{23}$) is less sensitive to the $\mu\tau$($e\tau$) final state. Smaller $\kappa$ relatively enhances the diagram (b) thus the neutrino mixing parameter becomes sensitive. Given the fact that the observed mixing parameters are almost compatible between normal and inverted ordering of the neutrino mass hierarchy while only $\delta_{CP}$ distinguishes the mass ordering, the difference of the production cross sections indicates the dependence of the $\delta_{CP}$ parameter. At smaller $\kappa$, for instance, $\kappa_{23}$ $=$ 0.01, about 30\% difference could be observed between normal and inverted ordering.

\begin{widetext}
\begin{center}
\begin{table}[tbhp]
\scalebox{0.95}{
\begin{tabular}{l|ccc|ccc} \hline \hline
Parameters & \multicolumn{3}{c|}{Normal ordering} & \multicolumn{3}{c}{Inverted ordering} \\
($m_{A}$, $\mathrm{tan}\beta$, $\kappa_{13}$, $\kappa_{23}$) & $\mu e$ [fb] & $\mu\tau$ [fb] & $e\tau$ [fb] & $\mu e$ [fb] & $\mu\tau$ [fb] & $e\tau$ [fb] \\ \hline
(500GeV, 40, 0.1, 0.1) & 2.258(6)$\times 10^{-1}$ 
                       & 1.862(3)$\times 10^{-1}$
                       & 2.100(5)$\times 10^{-1}$
                       & 2.30(1)$\times 10^{-1}$
                       & 1.831(9)$\times 10^{-1}$
                       & 2.09(1)$\times 10^{-1}$ \\
(800GeV, 40, 0.1, 0.1) & 2.272(8)$\times 10^{-1}$
                       & 1.389(7)$\times 10^{-1}$
                       & 1.612(3)$\times 10^{-1}$
                       & 2.29(1)$\times 10^{-1}$
                       & 1.346(8)$\times 10^{-1}$
                       & 1.62(1)$\times 10^{-1}$ \\
(1000GeV, 40, 0.1, 0.1) & 2.261(7)$\times 10^{-1}$
                        & 1.212(4)$\times 10^{-1}$
                        & 1.443(3)$\times 10^{-1}$
                        & 2.29(1)$\times 10^{-1}$
                        & 1.195(5)$\times 10^{-1}$
                        & 1.465(5)$\times 10^{-1}$ \\
(2000GeV, 40, 0.1, 0.1) & 2.255(6)$\times 10^{-1}$
                        & 9.70(4)$\times 10^{-2}$
                        & 1.198(3)$\times 10^{-1}$
                        & 2.29(1)$\times 10^{-1}$
                        & 9.45(4)$\times 10^{-2}$
                        & 1.224(4)$\times 10^{-1}$ \\
(5000GeV, 40, 0.1, 0.1) & 2.264(7)$\times 10^{-1}$
                        & 9.36(2)$\times 10^{-2}$
                        & 1.144(8)$\times 10^{-1}$
                        & 2.27(1)$\times 10^{-1}$
                        & 9.11(6)$\times 10^{-2}$
                        & 1.189(5)$\times 10^{-1}$ \\ \hline
(1000GeV, 10, 0.1, 0.1) & 4.34(2)$\times 10^{-6}$
                        & 2.54(1)$\times 10^{-4}$
                        & 2.60(1)$\times 10^{-4}$
                        & 4.22(2)$\times 10^{-6}$
                        & 2.55(1)$\times 10^{-4}$
                        & 2.58(1)$\times 10^{-4}$ \\
(1000GeV, 20, 0.1, 0.1) & 9.17(8)$\times 10^{-4}$
                        & 3.92(1)$\times 10^{-3}$
                        & 4.24(5)$\times 10^{-3}$
                        & 9.21(6)$\times 10^{-4}$
                        & 3.95(1)$\times 10^{-3}$
                        & 4.20(2)$\times 10^{-3}$ \\
(1000GeV, 30, 0.1, 0.1) & 2.27(1)$\times 10^{-2}$
                        & 2.33(1)$\times 10^{-2}$
                        & 2.64(1)$\times 10^{-2}$
                        & 2.33(2)$\times 10^{-2}$
                        & 2.29(1)$\times 10^{-2}$
                        & 2.70(1)$\times 10^{-2}$ \\ \hline
(2000GeV, 40, 0.01, 0.1) & 2.40(3)$\times 10^{-2}$
                         & 5.10(3)$\times 10^{-2}$
                         & 3.55(1)$\times 10^{-3}$
                         & 2.69(1)$\times 10^{-2}$
                         & 6.29(2)$\times 10^{-2}$
                         & 4.11(2)$\times 10^{-3}$ \\
(2000GeV, 40, 0.02, 0.1) & 2.98(2)$\times 10^{-2}$
                         & 5.44(3)$\times 10^{-2}$
                         & 6.97(2)$\times 10^{-3}$
                         & 3.22(1)$\times 10^{-2}$
                         & 6.56(7)$\times 10^{-2}$
                         & 8.32(3)$\times 10^{-3}$ \\
(2000GeV, 40, 0.05, 0.1) & 6.91(4)$\times 10^{-2}$
                         & 6.92(5)$\times 10^{-2}$
                         & 2.85(1)$\times 10^{-2}$
                         & 7.31(3)$\times 10^{-2}$
                         & 7.63(4)$\times 10^{-2}$
                         & 3.23(1)$\times 10^{-2}$ \\
(2000GeV, 40, 0.1, 0.01) & 2.54(1)$\times 10^{-3}$
                         & 4.29(4)$\times 10^{-4}$
                         & 4.04(8)$\times 10^{-2}$
                         & 2.69(2)$\times 10^{-3}$
                         & 3.30(4)$\times 10^{-4}$
                         & 2.90(3)$\times 10^{-2}$ \\
(2000GeV, 40, 0.1, 0.02) & 9.44(3)$\times 10^{-3}$
                         & 1.77(3)$\times 10^{-3}$
                         & 4.34(3)$\times 10^{-2}$
                         & 9.55(5)$\times 10^{-3}$
                         & 1.41(1)$\times 10^{-3}$
                         & 3.47(2)$\times 10^{-2}$ \\
(2000GeV, 40, 0.1, 0.05) & 5.68(4)$\times 10^{-2}$
                         & 1.34(3)$\times 10^{-2}$
                         & 6.73(3)$\times 10^{-2}$
                         & 5.4(1)$\times 10^{-2}$
                         & 1.24(1)$\times 10^{-2}$
                         & 5.73(2)$\times 10^{-2}$ \\ \hline\hline
\end{tabular}
}
\caption{
Summary of the production Cross section in different final states ($\mu e$, $\mu\tau$ and $e\tau$) with various input parameters, separated by the normal and inverted neutrino mass hierarchy. The first block is the cross sections by scanning of the $m_{A}$ with fixed $t_{\beta}$, $\kappa_{13}$ and $\kappa_{23}$ parameters. The second block is for the $\mathrm{tan}\beta$ scan, the third for the $\kappa_{13}$ and $\kappa_{23}$ scans. The unit is fb.
\label{tab:xsecsummary}
}
\end{table}
\end{center}
\end{widetext}

\section{V. Feasibility study for LHC}

The signal events are interfaced by Pythia \cite{Sjostrand:2006za} to adopt a parton shower in the hard-process and hadronize the color-charged quark and gluons radiated off from the colliding partons, and to simulate the other remnant interaction in the protons. The tau lepton is decayed by Tauola \cite{Jadach:1993hs}. The generated hadrons are reconstructed as a jet by a jet finder algorithm build-in Pythia with the similar experimental setup of the ATLAS/CMS calorimeter detectors \footnote{For simplicity, same calorimeter segment is used as the $\eta$ coverage of 4.9 with 0.025 fine cell granularity in $\phi$ and $\eta$ directions with 15\%$\sqrt{\mathrm{GeV}}$.}. Background processes are also generated. The $Z$ + $n$ jets ($n=0-4$) and diboson $WW$ + $m$ jets ($m=0,1,2$) processes are generated by Alpgen \cite{Mangano:2002ea}, where the order $\alpha_{em}$=4 $Z$ +2 jets processes are also included. The $t\bar{t}$ processes are generated by McAtNLO generator \cite{Frixione:2002ik} with NLO accuracy.

\begin{figure}[htbp]
\begin{center}
\begin{tabular}{c}
\includegraphics[width=7.5cm]{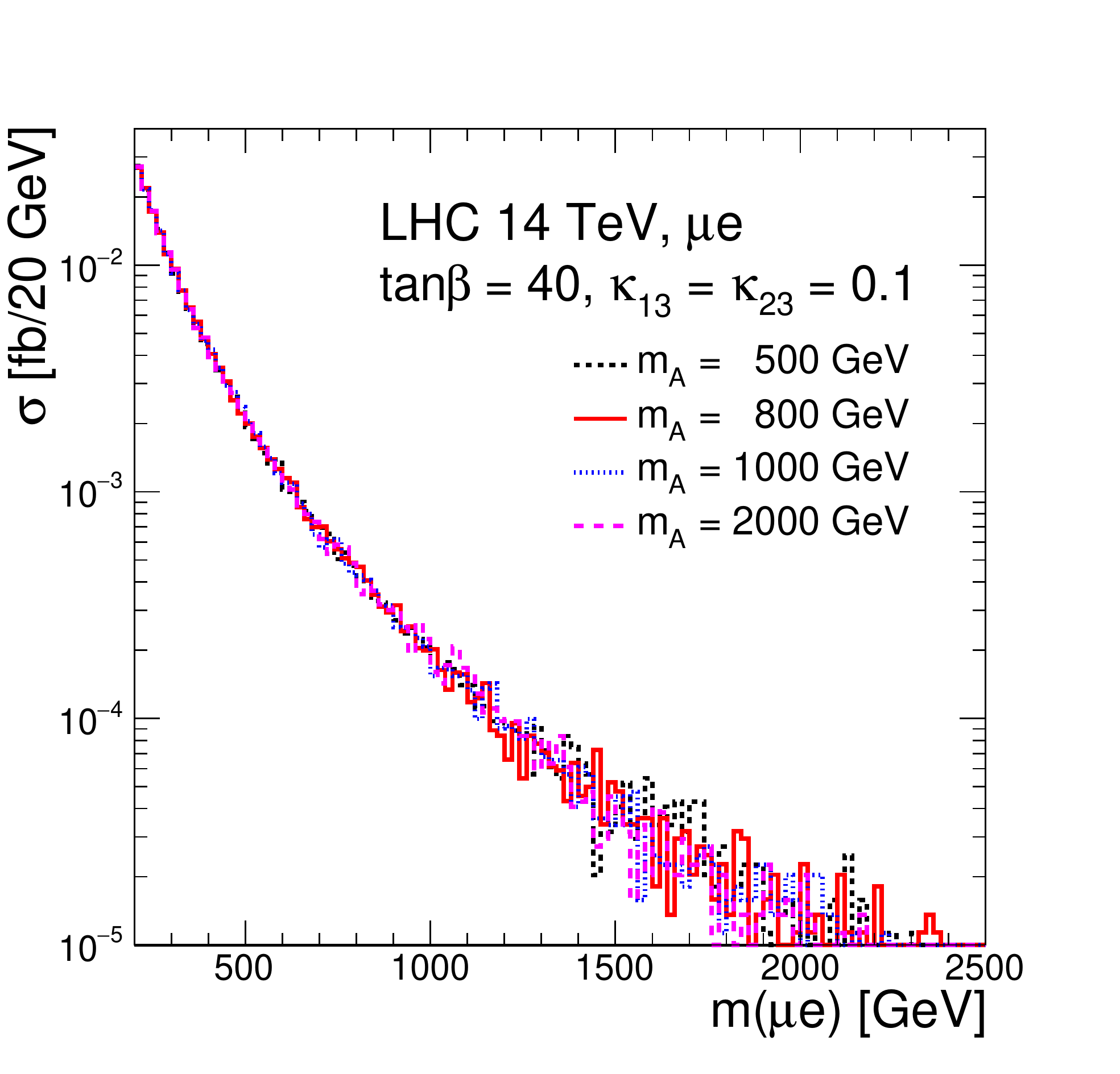} \\
(a) \\
\includegraphics[width=7.5cm]{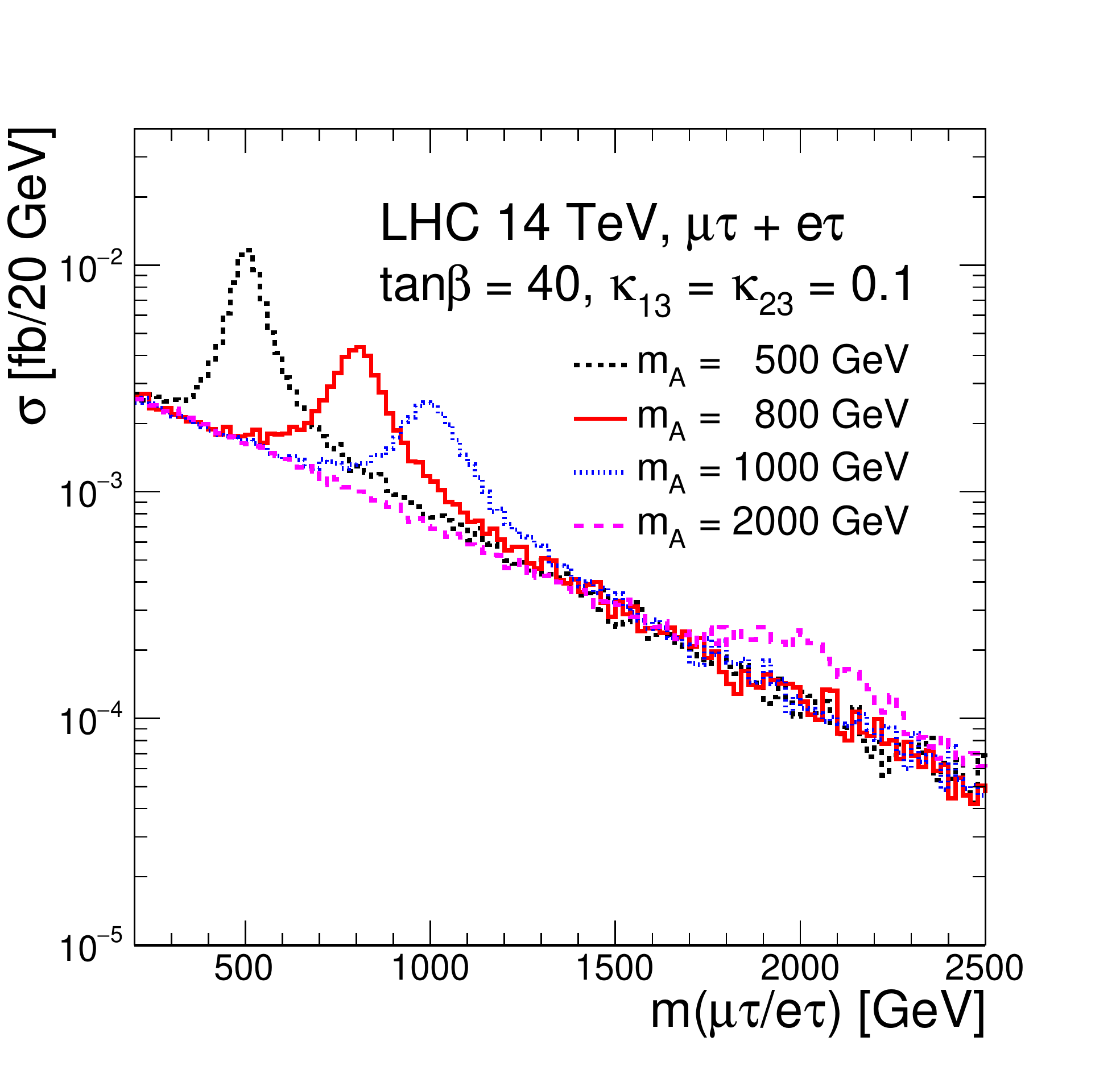} \\
(b) 
\end{tabular}
\caption{
Invariant mass distributions of lepton pair for with (a) $\mu e$, (b) $\mu\tau$ + $e\tau$ final states with $t_{\beta}$ = 40, $\kappa_{13}$ = $\kappa_{23}$ = 0.1.
}
\label{fig:lfvmassll}
\end{center}
\end{figure}

Three different final stats, $\mu e$, $\mu\tau$ and $e\tau$, are considered. The invariant masses of lepton pair are presented in Fig.\ref{fig:lfvmassll} (a) $\mu e$ and (b) $\mu\tau$ + $e\tau$ final states, respectively. As shown in Fig.\ref{fig:leading_diagrams}, the $\mu e$ final state has rather sharp falling while a mild slope with Higgs mass peaks in $\mu\tau$ + $e\tau$ final states due to the corresponding $s$-channel diagrams. At large $t_{\beta}$, although the $\mu e$ final state has larger production cross section.

An experimental feasibility is evaluated under those configurations. For simplicity, the muon and electron are assumed to be identified by 100\% efficiency within a fiducial volume of detector $|\eta|$ $<$ 2.5. No trigger efficiency is assumed. The jets are reconstructed with $p_{T}$ $>$ 25 GeV within $|\eta|$ $<$ 5.0. The hadronically decaying tau-lepton are only considered as the tau object ($\tau_{h}$) and assume 75 \% identification efficiency. The background rejection for quark and gluon jets misidentified is also taken into account as 3 \% for 1-prong and 0.4 \% for 3-prong. 
The $b$-jet is identified with 85 \% efficiency within the tracking volume of $|\eta|$ $<$ 2.5 and a light-flavour jet rejections 3.5 \%. 

The signal topology is two high energy leptons plus two jets. The flavor of leptons must be different with the opposite charges. Two jets are observed in opposite hemisphere with large invariant mass ($m_{jj}$ $>$ 500 GeV) and $\eta$ separation ($|\Delta\eta_{jj}|$ $>$ 5.0). There is no missing transverse energy ($E_{T}^{\mathrm{miss}}$ $<$ 10 GeV). The background processes are rejected by lepton ($\mu$, $e$ or $\tau_{h}$) $p_{T}$ $>$ 100 and 50 GeV, respectively. Since the neutrino is also associated in the $\tau_{h}$, the direction between $\tau_{h}$ and $E_{T}^{\mathrm{miss}}$ is used as $|\Delta\phi(\tau_{h},E_{T}^{\mathrm{miss}})|$ $<$ 0.05 instead of $E_{T}^{\mathrm{miss}}$ cut for $\mu\tau$ and $e\tau$ final state. After $b$-jet veto is applied to suppress the $t\bar{t}$ background, 13 events for $\mu e$, 11 for $\mu\tau$, and 13 for $e\tau$ are expected to be observed at the luminosity of 3000 fb$^{-1}$ against 53 background events for $\mu e$, 15 for $\mu\tau$, and 11 for $e\tau$ in the $m_{ll} > 500$ GeV region. 

\begin{figure}[htbp]
\begin{center}
\includegraphics[width=7.5cm]{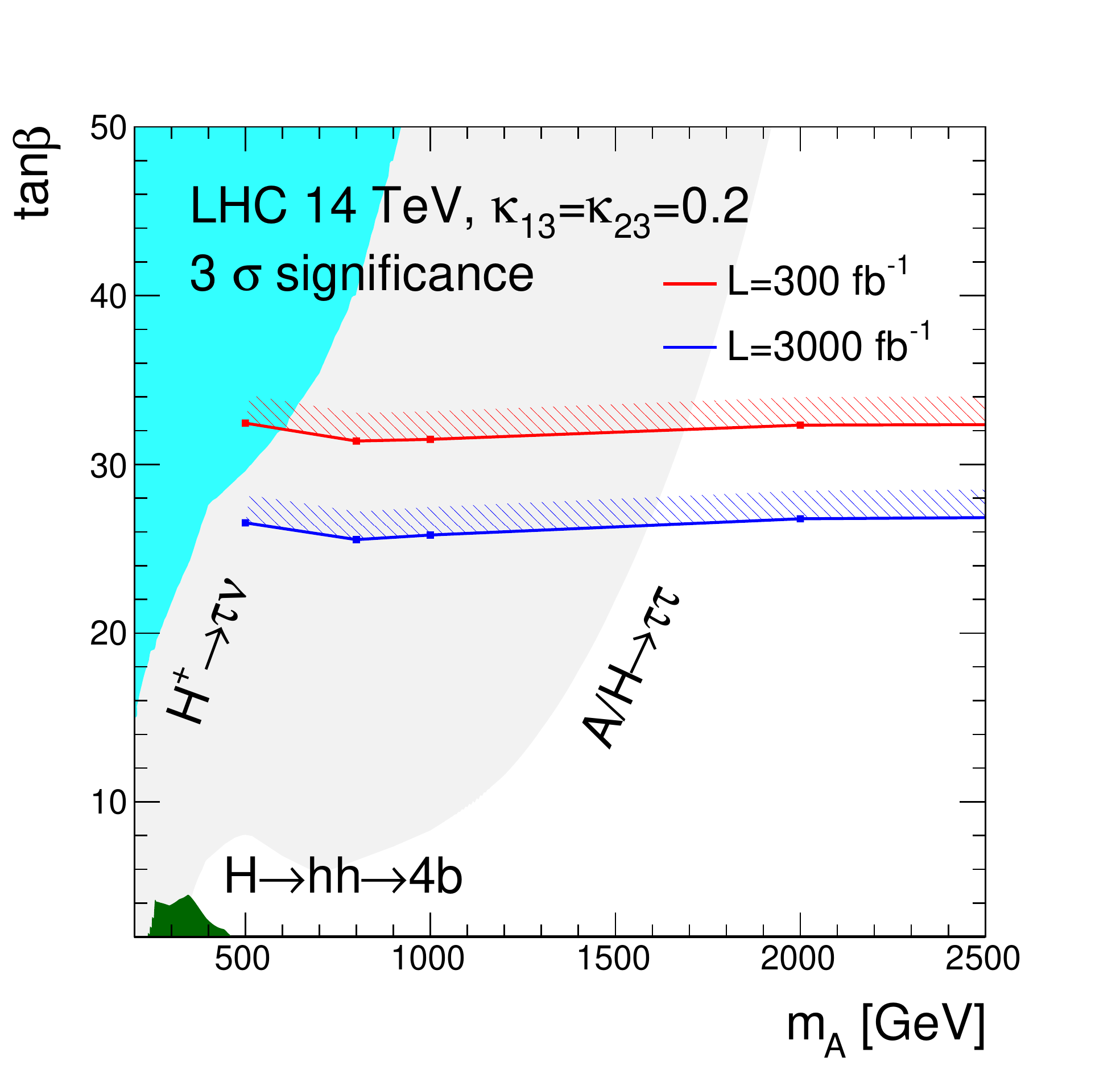}
\caption{
Expected 3$\sigma$ significance of $t_{\beta}$ as a function $m_{A}$ at $\kappa_{13}$ = $\kappa_{23}$ = 0.2, where the limits from three final states are combined. Two different luminosity scenario is presented. As reference, current limit from the ATLAS experiment is also shown.
}
\label{fig:lfvlimit}
\end{center}
\end{figure}

\begin{figure}[htbp]
\begin{center}
\includegraphics[width=7.5cm]{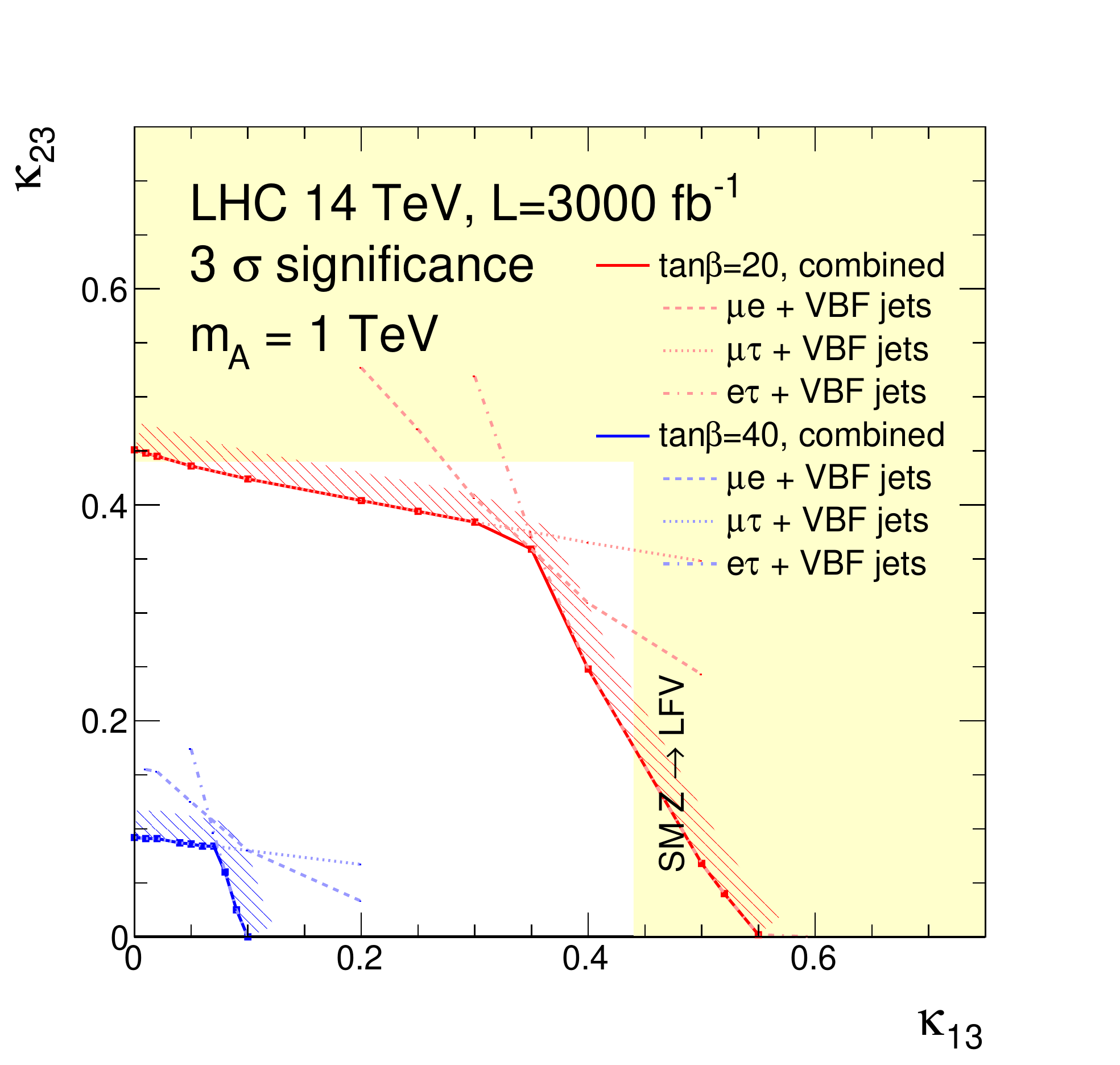}
\caption{
Exclusion plane as functions of $\kappa_{13}$ and $\kappa_{23}$ at $m_{A}$ = 1 TeV with 3000$fb^{-1}$ luminosity. Two different $t_{\beta}$ = 20 and 40 scenario are assumed. The $\kappa_{13}$, $\kappa_{23}$ $>$ 0.44 are already excluded in the tau decay measurements at $m_{A}$ = 1 TeV.
}
\label{fig:lfvlimitcoupling}
\end{center}
\end{figure}

The excess with 3$\sigma$ significance is evaluated as a function of $m_{A}$ for $\kappa_{13}$ = $\kappa_{23}$ = 0.2 by counting the number of signal and background events in Fig.\ref{fig:lfvlimit}, where the limits from three final states are combined. Two different luminosity scenarios with 300 and 3000 fb$^{-1}$ are considered. Current limits from the non-LFV MSSM Higgs boson searches \cite{mssmhiggsatlas} by the ATLAS experiment are also overlaid as reference, to see the sensitivity does not reach to higher mass region while such degradation is not observed in the LFV $t$-channel searches. With 300 fb$^{-1}$, the region of $t_\beta$ $>$ 30 is excluded for entire mass range. 
The limit of the $\kappa$ parameters are also scanned for given $t_{\beta}$. Figure \ref{fig:lfvlimitcoupling} presents the contour region of 3$\sigma$ exclusion limits in $\kappa_{13}$ and $\kappa_{23}$ plane for $m_{A}$ = 1 TeV and 3000 fb$^{-1}$ luminosity by single experiment. The $\mu\tau$ and $e\tau$ final states constrain the $\kappa_{23}$ and $\kappa_{13}$, respectively. while the $\mu e$ final state constrains both $\kappa_{13}$ and $\kappa_{23}$. With 3000 fb$^{-1}$ of data, the exclusion of $\kappa$ parameters reaches $\approx$ 0.1.

The $\kappa_{13}$, $\kappa_{23}$ $>$ 0.44 are already excluded in the tau decay measurements at $m_{A}$ = 1 TeV by the FCNC searches of $Z$ boson \cite{ATLAS:2020zlz}. The limits from the SM H decaying into LFV processes \cite{ATLAS:2019pmk, ATLAS:2019old} do not contribute in the constraint of the $\kappa_{13}$ and $\kappa_{23}$ due to the large suppression by the $\mathrm{cos}\alpha\sim0$ at large $m_{A}$ region. Meanwhile, the non-LFV neutral MSSM $H \rightarrow \tau\tau$ \cite{ATLAS:2020zms,CMS:2018rmh} could be re-interpreted from the observed cross section limit to constrain the $\kappa$ parameters. Their limits are about $\sigma(H/A \rightarrow \tau\tau) \lesssim$ 1-2 fb at $m_{A}=$1 TeV, which gives $\kappa_{13}(\kappa_{23}) \approx 0.3$ at $t_{\beta}$ $=$ 40.

\section{VI. Discussion and conclusion}

The LFV measurements at LHC should be compared with the constraints set by the measurements of LFV in the $\tau$ decay.
The most stringent constraints come from the rare decay of $\tau\to l \gamma$ and $\tau\to l \eta$. It is known that in THDM the constraints from the rare processes $\tau\to l \gamma$ are the strongest limit for heavier mass of $A$ due to the non-decoupling effect in the Barr-Zee diagrams \cite{Barr:1990vd,Chang:1993kw}. On the other hand, the decay width of $\tau\to l \gamma$ is strongly suppressed at $m_A\sim 700$ GeV due to cancellations at two loops, where the constraints from $\tau\to l \eta$ becomes most stringent. The constraints from $\tau\to l \gamma$ for generic Yukawa interaction including LFV have been discussed in Ref. \cite{Paradisi:2005tk}, and the constraints on each branching ratios are given as ${\rm Br}(\tau\to e\gamma)<3.3\times10^{-8}, {\rm Br}(\tau\to e\eta)<9.2\times10^{-8}, {\rm Br}(\tau\to \mu\gamma)<4.4\times10^{-8}, {\rm Br}(\tau\to \mu\eta)<6.5\times10^{-8}$ \cite{Zyla:2020zbs}.
For these channels, Belle II experiment is expected to improve a sensitivity by more than one order of magnitude when assuming an integrated luminosity of 50 ab$^{-1}$~\cite{Belle-II:2018jsg}.
These experimental bounds, in particular $\tau\to l_i\gamma$, can be translated into the constraints on $\kappa_{i3}$~\cite{Paradisi:2005tk}.
For $m_A \gtrsim m_W$,
\begin{eqnarray}
    \kappa_{i3} \lesssim 0.07\times\left(\frac{10}{t_\beta}\right)^2\sqrt{\frac{{\rm Br}_{\tau\to l_i\gamma}^{\rm exp}/{\rm Br}_{\tau\to l_i\nu\nu}^{\rm exp}}{2\times10^{-7}}}
\end{eqnarray}
can be obtained, where ${\rm Br}_{\tau\to l_i\gamma}^{\rm exp}$ is the experimental upper bound on the $\tau\to l_i\gamma$ channel, and ${\rm Br}_{\tau\to l_i\nu\nu}^{\rm exp}$ is the observed branching fraction.
Notice that this bound does not strongly depend on $m_A$ because of the non-decoupling nature of the Barr-Zee diagrams.

The VBF production mode in the heavy Higgs boson search with LFV will be a new physics process ever analyzed at LHC and provide new channels complementary to the LFV measurements in the $\tau$ decay. Especially, unlike a conventional decay mode of the Higgs boson to LFV, the $\mu e$ mode is enhanced by $t_{\beta}$ at high mass region. The dominant process through the 1-loop diagram is the $t$-channel production, thus the experimental search is accessible even higher mass region, which is not limited by the colliding energy. With 3000 fb$^{-1}$ of data, vast of the parameters space is explored at HL-LHC experiment. This will also serve as an input for the future collider experiments.

\subsection*{Acknowledgements}
\noindent
S.T. and K.K. acknowledge the support from the Ministry of Education, Culture, Sports, Science, and Technology (MEXT) of Japan, the Japan Society for the Promotion of Science (JSPS), the Grant-in-Aid for Scientific Research (C) 18K03685 and 19H01899. 

\bibliography{biblio} 

\end{document}